\documentstyle[12pt,psfig]{article}
\begin{document}

\hfill PUPT-1785

\hfill hep-th/9804066

\vspace{1.5in}

\begin{center}

{\large\bf Conjectures on }

{\large\bf (0,2) Mirror Symmetry }

\vspace{1in}

Eric Sharpe \\
Physics Department \\
Princeton University \\
Princeton, NJ  08544 \\
{\tt ersharpe@puhep1.princeton.edu }

\end{center}

In this paper we conjecture a reformulation of the monomial-divisor
mirror map for (2,2) mirror symmetry, valid at a boundary
of the moduli space, that is easily extended
to also include tangent bundle deformations -- an important step
towards understanding (0,2) mirror symmetry.
We check our conjecture in a few simple
cases, and thereby illustrate how to perform calculations
using a description of sheaves recently published by Knutson, Sharpe.

\begin{flushleft}
April 1998 
\end{flushleft}

\newpage

\section{Introduction}

Historically mirror symmetry has provided a fascinating interplay
between questions in string theory and algebraic geometry.
To review, mirror symmetry is simply a symmetry under which
two distinct Calabi-Yaus, say $X$ and $Y$, are both described
by the same conformal field theory -- string theory is unable
to distinguish them.
Mirror symmetry is to date not deeply understood, but a large number
of empirical results have been obtained.

There exists a potential generalization of mirror symmetry known
as (0,2) mirror symmetry, which has been developed much less far.
Whereas ordinary (so-called (2,2) ) mirror symmetry relates 
Calabi-Yaus $X$ and $Y$,
(0,2) mirror symmetry relates pairs $(X, {\cal E})$ and $(Y, {\cal F})$,
where ${\cal E}$ and ${\cal F}$ are torsion-free sheaves on $X$, $Y$,
respectively.  More precisely, there exists a single conformal
field theory which describes both pairs.  Ordinary mirror
symmetry is recovered in the special case ${\cal E} = TX$, ${\cal F}
= TY$.    

Ordinary mirror symmetry maps complex and Kahler moduli into one
another, so for three-folds
\begin{eqnarray*}
H^{(1,1)}(X) & \cong & H^{(2,1)}(Y) \\
H^{(2,1)}(X) & \cong & H^{(1,1)}(Y) 
\end{eqnarray*}
In at least the simplest possible examples of (0,2) mirror symmetry,
in which ${\cal E}$ and ${\cal F}$ are both deformations of the
tangent bundles of $X$, $Y$, 
\begin{eqnarray*}
\mbox{Ext}^1_X ( {\cal O}, {\cal E}) & \cong & \mbox{Ext}^1_Y (
{\cal F}, {\cal O} ) \\
\mbox{Ext}^1_X ( {\cal E}, {\cal O}) & \cong & \mbox{Ext}^1_Y (
{\cal O}, {\cal F})
\end{eqnarray*}
Potentially (0,2) mirror symmetry may relate the complex, Kahler,
and sheaf moduli of $(X, {\cal E})$ to the complex, Kahler,
and sheaf moduli of $(Y, {\cal F})$ in a highly intricate fashion,
though as yet no one knows.

Although ordinary mirror symmetry is not deeply understood, it has
been well-developed -- there are well-known methods to construct
mirror Calabi-Yaus, and precise relations between complex and Kahler
moduli have been developed.  No such statements are true of
(0,2) mirror symmetry\footnote{Except in the special limit in which
(0,2) mirror symmetry reduces to ordinary mirror symmetry, of course.},
although there has been a very limited amount of work on the subject
\cite{meallen,ralph1,ralph2,ralph3}.

The purpose of this paper is to begin to rectify this situation,
by describing steps towards a precise map between complex, Kahler,
and sheaf moduli in the special case that the sheaf over either
Calabi-Yau is a deformation of the tangent bundle.
 
More precisely, we make a precise conjecture for a reformulation
of the existing monomial-divisor mirror map for ordinary mirror symmetry
which easily generalizes to include sheaf deformations.
Our conjecture is only valid at large radius limits -- we do not
know how to improve its range of validity.   

We make a few simple tests of our conjecture, which allow us to 
demonstrate explicitly how to perform calculations using 
a description of sheaves
recently espoused in \cite{meallen}, which contained a review and
significant extension of results in \cite{kly1,kly2,kly3,kly4}.

\section{The Monomial-Divisor Mirror Map}

Let us briefly review the monomial-divisor mirror map
as described in for example \cite{agm}.

For any Calabi-Yau realized as a hypersurface\footnote{Generalizations
to complete intersections exist but are not relevant for this paper,
so they are omitted.} in a Fano toric variety, one can associate
two polytopes, call them ${\cal A}$ and ${\cal B}$.  

One polytope, call it ${\cal A}$, is the Newton polytope
of the complex structure -- it is the polytope containing all
points of the weight lattice $M$ which are associated to possible
monomials appearing in the hypersurface equation.  Put another
way, ${\cal A}$ is the (convex polytope) image of the moment
map associated to the $(S^1)^n \subset ( {\bf C}^{\times} )^n$ action 
on the toric
variety, with symplectic form defined by the anticanonical divisor.

The polytope ${\cal B}$ can be constructed as the polyhedron
generating the fan for the ambient toric variety, as cones over faces.
(Note that such a polyhedron can only exist when the toric
variety is projective.)
This polyhedron is also the ``polar polyhedron''
of ${\cal A}$ \cite[section 4.1]{fulton}:
\begin{displaymath}
{\cal B} \: = \: \{ (x_1, \cdots x_n) \in M_{\bf R} \, | \,
\sum_{i} x_i y_i \geq -1 \, \forall (y_1, \cdots y_n) \in {\cal A}
\, \}
\end{displaymath}

Mirror symmetry exchanges ${\cal A}$ and ${\cal B}$.
By examining how vertices of the polyhedra are mapped to one
another, we recover the monomial-divisor mirror map.
In particular, vertices of ${\cal A}$ correspond to monomials
and vertices of ${\cal B}$ correspond to toric divisors,
so we can see explicitly which monomials are mapped to which
divisors, and vice-versa.

Essentially the monomial-divisor mirror map acts by exchanging
convex polytope moment map images.  The vertices of such polytopes
correspond to fixed points of the (Hamiltonian) torus action;
by studying how fixed points are mapped to one another, we learn
how monomials and divisors are exchanged.

In what follows we will present a 
formally distinct but (hopefully) equivalent formulation of the
monomial-divisor mirror map, which directly exchanges specific
components of sheaf cohomology groups.

\section{Reformulation of Monomial-Divisor Mirror Map}

How might we extend the monomial-divisor mirror map
to include sheaf deformations?  In order to do so,
we shall work near large complex structure limit points.
In particular, not just any large complex structure limit
points, but those in which the tangent bundle of the Calabi-Yau
is stably equivalent to the restriction of an ``equivariant''
sheaf on the ambient toric variety.

\subsection{Equivariance}

What does it mean for a sheaf on a toric variety to be equivariant?
All toric varieties 
have a natural action
of $( {\bf C}^{\times} )^{n}$.  This action on the toric
variety defines an action on any moduli space of sheaves:
for any sheaf ${\cal E}$ and any $t \in ( {\bf C}^{\times} )^{n}$,
we can take ${\cal E} \mapsto t^{*} {\cal E}$.
Typically ${\cal E} \neq t^{*} {\cal E}$, but in the special
case that ${\cal E} = t^* {\cal E}$ for all $t \in ( {\bf C}^{\times} )^n$,
we say that ${\cal E}$ is equivariant\footnote{In fact,
it was noted in \cite{meallen} that this definition of equivariant
sheaves is not quite
correct technically, but for the purposes of this article it
will suffice.} (with respect to the
algebraic torus).

Equivariant sheaves have a number of nice properties.
For example, sheaf cohomology groups and global Ext groups
of equivariant sheaves on toric varieties have a canonical
decomposition, known as an isotypic decomposition,
by elements of the weight lattice $M$ of the algebraic torus.
More precisely, if ${\cal E}$ and ${\cal F}$ are
equivariant, then
\begin{displaymath}
H^p ( {\cal E} ) \: = \: \bigoplus_{\chi} H^p ( {\cal E} )_{\chi} 
\end{displaymath}
\begin{displaymath}
\mbox{Ext}^p ( {\cal E}, {\cal F} ) \: = \:
\bigoplus_{\chi} \mbox{Ext}^p ( {\cal E}, {\cal F} )_{\chi}
\end{displaymath}

We should note that when authors speak of equivariant sheaves,
they sometimes implicitly assume a specific choice of ``equivariant 
structure'' has been made.  An equivariant structure is simply
a precise choice of action of the algebraic torus on the sheaf;
it is not sufficient to know the fact that the algebraic torus
maps the sheaf back into itself, one must also know precisely
how the algebraic torus maps the sheaf into itself.

Under what circumstances is the tangent bundle of a Calabi-Yau
stably equivalent to the restriction of an equivariant sheaf?
This will happen when the hypersurface defining the Calabi-Yau
is a monomial.  Such a Calabi-Yau is highly degenerate, and
for technical reasons degenerate Calabi-Yaus are often excised from 
moduli spaces
of complex structures.  The degenerate Calabi-Yaus we shall
consider, however, still exist on the moduli space. 
(For more information
on complex structure moduli spaces, see appendix~\ref{compmodspace}.)
We will refer to
(degenerate) Calabi-Yaus whose tangent bundles are stably
equivalent to the restriction of an equivariant sheaf as
``equivariant Calabi-Yaus''.

If we describe the ambient toric variety in terms of homogeneous
coordinates and ${\bf C}^{\times}$ actions, then the product
of all homogeneous coordinates is a possible monomial appearing
in a Calabi-Yau hypersurface equation, and by itself represents a 
degenerate Calabi-Yau that exists on the moduli space.
Better, in fact:  
it is usually a large complex structure limit \cite{vafaleung,aspinwallpriv}.

Let us consider a specific example.  Let $x_1, \cdots x_5$
represent homogeneous coordinates on ${\bf P}^{4}$ and
${\bf P}^{4}/({\bf Z}_5)^3$, then in both ambient spaces
the hypersurface
\begin{displaymath}
x_1 x_2 x_3 x_4 x_5 \: = \: 0
\end{displaymath}
is a degenerate Calabi-Yau.  Moreover, as is well-known,
this hypersurface in ${\bf P}^{4}/({\bf Z}_5)^3$ is precisely
the large complex structure limit.
Another example \cite{candI} is the family of degree 8 hypersurfaces
in ${\bf P}^{4}_{1,1,2,2,2}/{\bf Z}_4^3$ given by
\begin{displaymath}
4 \psi x_1 x_2 x_3 x_4 x_5 \: + \: \phi x_1^4 x_2^4 \: = \: 0
\end{displaymath}
where $x_1, \cdots x_5$ are the homogeneous coordinates.
This family of Calabi-Yaus represents the large complex structure
limits of the mirror to degree 8 hypersurfaces in ${\bf P}^4_{1,1,2,2,2}$.
When either $\psi = 0$ or $\phi = 0$, we recover an equivariant
Calabi-Yau.

As this point is quite important, we shall repeat it.
In the ``classical'' limit of mirror symmetry, in which
all worldsheet instanton corrections are suppressed in
both mirror Calabi-Yaus, both Calabi-Yaus are degenerate
and in limits
have tangent bundles stably equivalent to the restriction
of an equivariant sheaf.  
Thus, mirror Calabi-Yaus exist which are both equivariant,
and such Calabi-Yaus correspond to a ``classical'' limit of mirror symmetry.

On an equivariant Calabi-Yau, 
sheaf cohomology groups and Ext groups inherit nontrivial properties
from the ambient toric variety.  In particular, in the case
that a sheaf on an equivariant Calabi-Yau $X$ is 
the restriction of
of an equivariant sheaf ${\cal E}$ on an ambient toric variety $Y$, the sheaf
cohomology groups $H^p(X, {\cal E}|_X )$ inherit the isotypic decomposition
of sheaf cohomology groups on the ambient space.
In particular, recall that the restriction of ${\cal E}$ to $X$
is defined by 
\begin{displaymath}
0 \rightarrow {\cal E} \otimes {\cal O}(K) \rightarrow
{\cal E} \rightarrow {\cal E} \otimes {\cal O}_{X} 
\rightarrow 0
\end{displaymath}
where ${\cal O}_X$ is a skyscraper sheaf with support on the
Calabi-Yau $X$.
Now, both ${\cal E} \otimes {\cal O}(K)$ and ${\cal E}$
are equivariant, and moreover the map
${\cal E} \otimes {\cal O}(K) \rightarrow {\cal E}$
is equivariant when $X$ is an equivariant Calabi-Yau, 
so clearly the sheaf cohomology groups $H^p( X, {\cal E}|_X )$
inherit the isotypic decomposition of sheaf cohomology
groups on the ambient toric variety $Y$.  Put another way,
in the case of an equivariant Calabi-Yau, the skyscraper
sheaf ${\cal O}_{X}$ is an equivariant torsionful sheaf,
so ${\cal E} \otimes {\cal O}_X$ is equivariant and
$H^p( Y , {\cal E} \otimes {\cal O}_X ) = H^p( X, {\cal E}|_X )$
has an isotypic decomposition.

\subsection{The Conjectured Reformulation}

Now, let us re-examine the monomial-divisor mirror map
in the limit of large complex structure and large radius,
and more specifically, in a limit in which both Calabi-Yaus
have tangent sheaves which are the restriction of an equivariant
sheaf.  In this limit, we conjecture the monomial-divisor mirror map
exchanges isotypic components of sheaf cohomology.
More precisely, if $X$, $Y$ are a mirror pair of Calabi-Yaus,
then in this limit, for all $\chi$ in the weight lattice
of the algebraic torus $M$, we claim
\begin{displaymath}
H^1 ( X, TX )_{\chi} \: \cong \: H^1 ( Y, T^* Y )_{\chi} 
\end{displaymath}
\begin{displaymath}
H^1 ( X, T^* X )_{\chi} \: \cong \: H^1 ( Y, TY )_{\chi}
\end{displaymath}
We conjecture that this is precisely the statement of the monomial-divisor
mirror map in the equivariant limit.

In fact, we have been slightly sloppy.  The weight lattices
of the algebraic tori underlying the ambient toric varieties
in which $X$ and $Y$ are embedded are not canonically isomorphic.
The more nearly correct statement of the conjecture is that
there exists an isomorphism of the weight lattices such that
components of sheaf cohomology groups are exchanged as above.

Not only have we been sloppy, we have also been somewhat naive.
Equivariant Calabi-Yaus are highly singular, and in the limit
of complex structure in which a Calabi-Yau becomes equivariant,
the sheaf cohomology groups may jump -- they may lose or gain
elements.  In particular, we shall see in specific examples that
it is quite common for the sheaf cohomology groups of an equivariant
Calabi-Yau to not quite have the same dimensions as the sheaf cohomology
groups of a smooth Calabi-Yau.  For example, we will find that
whereas a smooth elliptic curve $E$ has $h^1(E, TE) = 1$,
for a degenerate $E$ we compute $h^1(E, TE) = 0$. 
Unfortunately this jumping phenomenon makes our conjecture somewhat
less palatable.
In order for our conjecture to be genuinely useful, one would need
to extend it away from equivariant limits.

\subsection{(0,2) Generalization}

How can we extend the monomial-divisor mirror map,
as reformulated above,
to include sheaf deformations?
When phrased in the language above, it is straightforward.
Deformations of a sheaf ${\cal E}$ are parametrized by
elements of (global) $\mbox{Ext}^1 ( {\cal E}, {\cal E} )$,
so the answer should be relatively clear.
In the special case that the gauge sheaves ${\cal E}$, ${\cal F}$ 
on $X$, $Y$, respectively are
both 
deformations of the tangent sheaves, and also happen
to be the restriction of equivariant sheaves, we can 
hypothesize that (0,2) mirror symmetry relates
\begin{displaymath}
\mbox{Ext}^1_{X} ( {\cal E}, {\cal E} )_{\chi} \: \cong \:
\mbox{Ext}^1_{Y} ( {\cal F}, {\cal F} )_{\chi} 
\end{displaymath}

How can such a result be interpreted?
In general, isotypic components of $\mbox{Ext}^1({\cal E}, {\cal E})$
associated with character 0 describe deformations that
preserve equivariance, whereas those with nonzero character
destroy equivariance.  If our conjecture is correct,
then it implies that deformations preserving equivariance
are mirror to deformations also preserving equivariance,
and also relates specific non-equivariance-preserving
deformations (at least in limits where worldsheet instanton
effects are small).

So far we have given a conjecture for a reformulation of the
monomial-divisor mirror map that works only in the limit in which
both Calabi-Yaus are at large radius and large complex structure.
How might one attempt to extend our ansatz away from this limit?
Consider the case that one of the Calabi-Yaus is deformed to
generic complex structure, but held at large radius (in fact at 
the same large radius limit).
In this case, the algebraic torus defining the ambient toric
variety no longer has a well-defined action on the sheaf
cohomology groups -- in fact, its action changes the complex
structure of the Calabi-Yau.  However, since the Calabi-Yau is
at large radius, its mirror is at large complex structure (and in fact 
still equivariant) -- thus, the algebraic torus defining the ambient
toric variety of the mirror Calabi-Yau does have an action!

Put another way, if we deform one of our equivariant Calabi-Yaus
to arbitrary complex structure but hold the Kahler moduli at the
same large radius point, then its sheaf cohomology still has
an isotypic decomposition -- but under the algebraic torus
defining the ambient toric variety of the mirror !
Similarly, if we hold the complex structure fixed but vary
the Kahler moduli, then the isotypic decomposition will still
be well-defined, although cup and Yoneda products will receive
worldsheet instanton corrections.  

Unfortunately, because the sheaf cohomology groups jump in
degenerations, there is no reason why the mirror isotypic
decomposition of sheaf cohomology of a smooth Calabi-Yau 
described above should coincide
with the isotypic decomposition of its degenerate mirror.
Our calculations of sheaf cohomology of degenerations are
somewhat naive -- perhaps we have missed subtleties which give
better-behaved results.
In any event, we will not see the isotypic decomposition under
a mirror algebraic torus explicitly in this paper.
If it were possible to correct  our ansatz for these cases, then 
one might wonder whether it can be extended farther -- to
the case when both the complex and Kahler moduli are generic.
We do not have any comments on this case, except to speculate
that some sort of derived categories argument may prove crucial.

In order to try to test our conjecture, we shall compute sheaf
cohomology groups and $\mbox{Ext}$ groups on degenerate
elliptic curves.  We shall not be able to check
whether any matching suggested by the isotypic decomposition
agrees with the usual monomial-divisor mirror map, we shall
only be able to see whether (judging by dimensions of components)
there exist matchings suggested by isotypic decompositions.

\subsection{Discrete R-anomalies}

There is a potential danger in our approach which we have glossed
over so far.  We shall be studying sheaves on toric varieties
that restrict to a sheaf stably equivalent to the tangent bundle
of the Calabi-Yau (for an elliptic curve $E$, sheaves ${\cal E}$ 
such that ${\cal E} |_E = TE \oplus {\cal O}$) and it was 
noted in \cite{distRanom} that linear sigma models describing
such situations suffer from certain poorly understood effects.

However we feel this is not a problem in the present case.
First, we are not trying to construct conformal field theories
directly but rather are merely studying a mathematical construction
designed to shed light on the monomial-divisor mirror map.

Second, we should point out the existence of discrete R-anomalies
at Landau-Ginzburg points does not reflect instability in a gauge
sheaf, despite claims in the existing literature.  On a Calabi-Yau
$X$, the sheaf $TX \oplus {\cal O}$ is not unstable but rather
properly semistable\footnote{More generally \cite{friedprivate}, the direct
sum of two Mumford-Takemoto semistable bundles with the same slope
is again Mumford-Takemoto semistable.}, and the $D$-term constraint
in the low-energy supergravity allows not only stable sheaves
but also split properly semistable sheaves \cite{meallen}.  
Properly semistable
sheaves are grouped\footnote{More precisely, points on a moduli
space of sheaves that are properly semistable do not correspond
to unique semistable sheaves, but rather to $S$-equivalence classes
of properly semistable sheaves.  Points that are stable do correspond
to unique stable sheaves -- $S$-equivalence classes are a phenomenon
arising only for properly semistable objects.  } in $S$-equivalence 
classes, and each
$S$-equivalence class contains a unique split representative
\cite[p.23]{huybrechtslehn}.  Thus, the discrete R-anomalies described
in \cite{distRanom} do not have anything to do with stability per se,
but rather quite possibly reflect the fact that the renormalization
group flow is extremely subtle to understand because of the presence
of a properly semistable sheaf.

\section{Elliptic Curves}

In this section we will perform consistency checks on
our conjecture 
by explicitly calculating relevant sheaf cohomology groups on
the degenerate elliptic curve in ${\bf P}^2$ and
on the mirror curve in ${\bf P}^2 / {\bf Z}_3$.

This particular case is extremely simple:  there is only one
complex, Kahler, and bundle deformation, so our conjecture
automatically holds true in this case.  In more complicated
cases we can check that dimensions of isotypic components of
sheaf cohomology groups match, but unfortunately nowhere 
in this paper (except in the trivial case of elliptic curves)
will we be able to determine whether the monomial-divisor mirror
map actually exchanges isotypic components.  All we can do
is check that dimensions of isotypic components match -- a consistency
test, no more.

As the techniques needed to perform our calculations are 
new to the physics literature, we shall work through the
calculations in detail.

\subsection{Elliptic Curve in ${\bf P}^2$ }

\begin{figure}
\centerline{\psfig{file=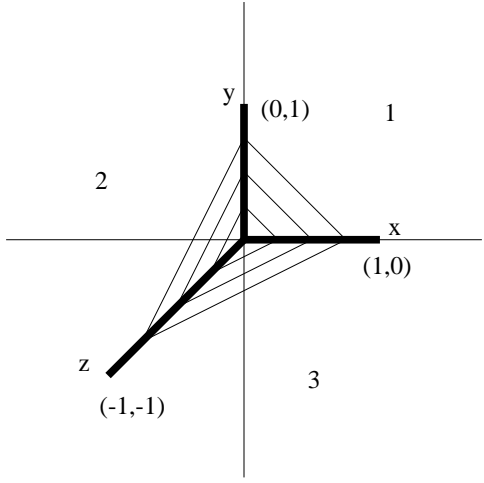,width=2.5in}}
\caption{\label{p2} A fan for the toric variety ${\bf P}^2$.}
\end{figure}

A fan describing ${\bf P}^2$ as a toric variety is shown in 
figure~\ref{p2}.

We will be interested in the degenerate elliptic
curve $E$ defined by
\begin{displaymath}
xyz \: = \: 0
\end{displaymath}
The tangent bundle of E is stably equivalent
to the restriction of an equivariant bundle, say ${\cal E}$, 
on ${\bf P}^2$.  (In other words, ${\cal E} |_{E} = TE \oplus
{\cal O}$.)
In the notation of \cite{meallen}, the bundle ${\cal E}$ on ${\bf P}^2$
is defined by the filtrations
\begin{displaymath}
({\cal E})^{\alpha}(i) \: = \: \left\{ \begin{array} {ll}
                               {\bf C}^2 &  i < 0 \\
                               \alpha   & i = 0,1 \\
                               0        & i > 1 
                               \end{array} \right.
\end{displaymath}
(To check that this bundle restricts on $E$ to become stably equivalent
to $TE$, i.e., that it is the same as the bundle on ${\bf P}^2$ defined by the
short exact sequence 
\begin{displaymath}
0 \: \rightarrow \: {\cal E} \: \rightarrow \: \bigoplus_{1}^{3} {\cal O}(1)
\: \rightarrow \: {\cal O}(3) \: \rightarrow \: 0
\end{displaymath}
note that both bundles have $c_1 = 0$ and $c_2 = 3$.)
In order to calculate the sheaf cohomology of $TE$, 
we shall calculate the sheaf cohomology of ${\cal E}$ and
${\cal E} \otimes {\cal O}(K)$ and use relations derived
from the short exact sequence
\begin{displaymath}
0 \: \rightarrow \: {\cal E} \otimes {\cal O}(K) \: \rightarrow \:
{\cal E} \: \rightarrow \: {\cal E} \otimes {\cal O}_{E} \:
\rightarrow \: 0
\end{displaymath}

The modules associated to the toric divisors have the forms
indicated below.  The module associated to $\{x = 0\}$,
call it $({\cal E})^x$, has the form
\begin{center}
\begin{tabular}{c|cccccc}
 & -3 & -2 & -1 & 0 & 1 & 2 \\ \hline
 &  & & & $\vdots$ & & \\
1 & & 0 & $(x)$ & $(x)$ & ${\bf C}^2$ & \\
0 & $\cdots$ & 0 & $(x)$ & $(x)$ & ${\bf C}^2$ & $\cdots$ \\
-1 & & 0 & $(x)$ & $(x)$ & ${\bf C}^2$ & \\
 & & & & $\vdots$ & & \\
\end{tabular}
\end{center}
The complex one-dimensional subspace of ${\bf C}^2$ generated
by $x = (1,0)$ is denoted by $(x)$.  Similarly, the module
associated to $\{y = 0\}$, call it $({\cal E})^y$, has the form
\begin{center}
\begin{tabular}{c|ccccc}
 & -2 & -1 & 0 & 1 & 2 \\ \hline
 &    &    & $\vdots$ & & \\
1 &   & ${\bf C}^2$ & ${\bf C}^2$ & ${\bf C}^2$ & \\
0 & $\cdots$ & $(y)$ & $(y)$ & $(y)$ & $\cdots$ \\
-1 &  & $(y)$ & $(y)$ & $(y)$ & \\
-2 &  & 0 & 0 & 0 & \\
   &  &  & $\vdots$ & & \\
\end{tabular}
\end{center}
and finally the module $({\cal E})^z$ associated to the
divisor $\{z = 0\}$ has the form
\begin{center}
\begin{tabular}{c|ccccc}
 &  & -1 & 0 & 1 & \\ \hline
 &  &    & $\vdots$ & & \\
3 & & 0 & 0 & 0 & \\
2 & & $(z)$ & 0 & 0 & \\
1 & & $(z)$ & $(z)$ & 0 & \\
0 & $\cdots$ & ${\bf C}^2$ & $(z)$ & $(z)$ & $\cdots$ \\
-1 & & ${\bf C}^2$ & ${\bf C}^2$ & $(z)$ & \\
-2 & & ${\bf C}^2$ & ${\bf C}^2$ & ${\bf C}^2$ & \\
  & & & $\vdots$ & & \\
\end{tabular}
\end{center}

\subsubsection{$H^1(E, {\cal E} )$}

Now, given the fact that
\begin{displaymath}
H^0( {\cal E} )_{\chi} \: = \: \bigcap_{\alpha \in | \Sigma |}
({\cal E})^{\alpha}(\chi)
\end{displaymath}
for reflexive sheaves ${\cal E}$,
and that $(x) \cap (y) = 0$, we can easily show that
\begin{displaymath}
H^0 ({\bf P}^2, {\cal E})_{\chi} \: = \: 0 \: \forall \chi
\end{displaymath}

In general, for equivariant bundles ${\cal E}$ on a smooth toric surface
\begin{displaymath}
H^2 ( {\cal E} )_{\chi} \: = \: \frac{V}{ \sum_{\alpha \in | \Sigma |}
({\cal E})^{\alpha}(\chi) }
\end{displaymath}
What is $V$?  Over the open $T$-orbit any equivariant torsion-free sheaf
is a trivial vector bundle; $V$ is the fiber of that bundle.
(In the present case, $V$ = ${\bf C}^2$.)
For the case at hand we can quickly show
\begin{displaymath}
H^2( {\bf P}^2, {\cal E} )_{\chi} \: = \: 0 \: \forall \chi
\end{displaymath}

We can then use
\begin{displaymath}
\sum_{i} (-)^i \mbox{dim } H^i ({\cal E})_{\chi} \: = \:
\sum_{\sigma \in \Sigma} (-)^{codim \: \sigma} \mbox{dim }
({\cal E})^{\sigma}(\chi)
\end{displaymath}
(where the origin of the fan is included as a (zero-dimensional) cone,
with associated module $V$)
to show that, for the case at hand,
\begin{displaymath}
H^1({\bf P}^2, {\cal E})_{\chi} \: = \: \left\{ \begin{array}{ll}
                        {\bf C} & \chi = 0 \\
                        0       & \mbox{otherwise}
                        \end{array} \right.
\end{displaymath}

The equivariant bundle ${\cal E} \otimes {\cal O}(K)$,
for the ``symmetric'' equivariant structure on $K$
(namely, $-K = \sum_{\alpha \in | \Sigma |} D_{\alpha}$)
is defined by the filtrations
\begin{displaymath}
( {\cal E} \otimes {\cal O}(K) )^{\alpha}(i) \: = \: \left\{ \begin{array}{ll}
                    {\bf C}^2 & i < -1 \\
                    \alpha    & i = -1, 0 \\
                    0         & i > 0
                    \end{array}  \right.
\end{displaymath}
and proceeding as before, one can quickly show
\begin{eqnarray*}
H^0 ( {\bf P}^2, {\cal E} \otimes {\cal O}(K) )_{\chi} & = & 0 \: \forall
\chi \\
H^2 ( {\bf P}^2, {\cal E} \otimes {\cal O}(K) )_{\chi} & = & 0 \: \forall
\chi \\
H^1 ( {\bf P}^2, {\cal E} \otimes {\cal O}(K) )_{\chi} & = & 
\left\{ \begin{array}{ll}
      {\bf C} & \chi = 0 \\
      0       & \mbox{otherwise}
      \end{array}  \right.
\end{eqnarray*}

Putting this together it is easy to see that
\begin{displaymath}
H^0(E, {\cal E})_{\chi} \: = \: H^1(E, {\cal E})_{\chi} \: = \: 0 \:
\forall \chi
\end{displaymath}
(For $\chi \neq 0$, the result is clear.  The result for
$\chi = 0$ must be obtained by a close examination of the exact
sequence.)

This result is an example of an unfortunate phenomenon mentioned
earlier -- in the singular complex structure limits describing
equivariant Calabi-Yaus, the sheaf cohomology groups may jump.

\subsubsection{$H^1(E, {\cal E}^{\vee})$}

How does one calculate the dual ${\cal E}^{\vee}$ of a 
sheaf ${\cal E}$ ?

More generally, given any two equivariant torsion-free sheaves
${\cal E}$ and ${\cal F}$, say, which over the open torus orbit
look like vector bundles with fiber $E$, $F$, respectively,
then to any cone $\sigma \in \Sigma$ we can define the 
module $\mbox{Hom}({\cal E},{\cal F})^{\sigma}$
as follows.  Each equivariant element 
$\mbox{Hom}({\cal E},{\cal F})^{\sigma}(\chi)$ is precisely a subspace
of $\mbox{Hom}(E,F)$.  Which subspace?  Well, any element of
$\mbox{Hom}(E,F)$ defines a map between equivariant elements of the modules
$({\cal E})^{\sigma}$ and $({\cal F})^{\sigma}$, and in particular
$\mbox{Hom}({\cal E}, {\cal F})^{\sigma}(\chi)$ is the subspace
of $\mbox{Hom}(E,F)$ that maps $({\cal E})^{\sigma}(\mu)$ into
$({\cal F})^{\sigma}(\mu + \chi)$ for all $\mu$.
(A general element of $\mbox{Hom}(E,F)$ will map
$({\cal E})^{\sigma}(\mu) \subseteq E$ into $F$ for any $\mu$, 
but not necessarily
into the subspace $({\cal F})^{\sigma}(\mu + \chi) \subseteq F$.)

How can we calculate ${\cal E}^{\vee}$ given ${\cal E}$ ?
Well, ${\cal E}^{\vee} = \mbox{Hom}({\cal E},{\cal O})$,
and in fact
it is straightforward to show that for any equivariant torsion-free
sheaf ${\cal E}$, the equivariant reflexive sheaf ${\cal E}^{\vee}$ is defined
by the filtrations
\begin{displaymath}
({\cal E}^{\vee})^{\alpha}(i) \: = \: \left( E / ({\cal E})^{\alpha}(1-i)
\right)^*
\end{displaymath}

In particular, for the case we are interested in,
\begin{displaymath}
({\cal E}^{\vee})^{\alpha}(i) \: = \: \left\{ \begin{array}{ll}
         ( {\bf C}^{2} )^* \cong {\bf C}^2 & i < 0 \\
         ( {\bf C}^2 / (\alpha) )^* = \{\omega \in ({\bf C}^2)^* |
            \langle \omega , \alpha \rangle = 0 \} & i = 0, 1 \\
         0 & i > 1
         \end{array} \right.
\end{displaymath}

In order to derive the sheaf cohomology of ${\cal E}^{\vee}$
and ${\cal E}^{\vee} \otimes {\cal O}(K)$, we can either proceed
directly as before, or we can use equivariant Serre duality,
which says that for any equivariant bundle ${\cal E}$ on an $n$-dimensional
projective toric variety $X$,
\begin{displaymath}
H^i ( X, {\cal E} )_{\chi} \: \cong \: H^{n-i} ( X, {\cal E}^{\vee} \otimes
{\cal O}(K) )_{ - \chi}^{\vee}
\end{displaymath}
where ${\cal O}(K)$ is assumed to have the ``symmetric'' equivariant
structure (namely, $K = - \sum_{\alpha \in | \Sigma | } D_{\alpha}$.)

In either event, one quickly derives that
\begin{displaymath}
H^0( E, {\cal E}^{\vee} )_{\chi} \: = \: H^1( E, {\cal E}^{\vee} )_{\chi}
\: = \: 0 \: \forall \chi
\end{displaymath}

\subsubsection{$\mbox{Ext}^1_{E}({\cal E}, {\cal E})$}

Using the methods outlined earlier, it is straightforward to compute
the filtrations defining the bundle $\mbox{Hom}({\cal E}, {\cal E})$.
Define
\begin{eqnarray*}
M & = & \mbox{Hom}({\bf C}^2, {\bf C}^2) \\
P^{\alpha} & = & \{ \omega \in M | \omega(\alpha) \subseteq (\alpha) \} \\
Q^{\alpha} & = & \{ \omega \in M | \mbox{im } \omega \subseteq (\alpha)
      \mbox{ and } \omega(\alpha) = 0 \} 
\end{eqnarray*}
then the filtrations defining $\mbox{Hom}({\cal E}, {\cal E})$ are
given by
\begin{displaymath}
\mbox{Hom}({\cal E},{\cal E})^{\alpha}(i) \: = \: \left\{ \begin{array}{ll}
             M   &  i < -1 \\
             P^{\alpha} & i = -1 , 0 \\
             Q^{\alpha} & i = 1 , 2 \\
             0   & i > 2 
             \end{array} \right.
\end{displaymath}

For bundles ${\cal E}$, $\mbox{Ext}^k({\cal E},{\cal E}) = H^k(
\mbox{Hom}({\cal E},{\cal E}))$; we shall use the notations interchangeably
in the rest of this subsection.

Using the same techniques as earlier it is now straightforward to
calculate
\begin{eqnarray*}
\mbox{Ext}^0_{{\bf P}^2}({\cal E},{\cal E})_{\chi} & = & 
   \left\{ \begin{array}{ll}
   {\bf C} & \chi = 0 \\
   0       & \mbox{otherwise} 
   \end{array} \right.   \\
\mbox{Ext}^2_{{\bf P}^2}({\cal E},{\cal E})_{\chi} & = & 0 \: \forall \chi 
\end{eqnarray*}
and $\mbox{Ext}^1_{{\bf P}^2}({\cal E},{\cal E})_{\chi}$ is given
by
\begin{center}
\begin{tabular}{c|cccccc}
 & -3 & -2 & -1 & 0 & 1 & 2  \\ \hline
2 & 0 & 0 & 0 & 0 & 0 & 0  \\
1 & 0 & ${\bf C}$ & ${\bf C}$ & ${\bf C}$ & ${\bf C}$ & 0 \\
0 & 0 & 0 & ${\bf C}$ & 0 & ${\bf C}$ & 0 \\
-1 & 0 & 0 & 0 & ${\bf C}$ & ${\bf C}$ & 0 \\
-2 & 0 & 0 & 0 & 0 & ${\bf C}$ & 0 \\
-3 & 0 & 0 & 0 & 0 & 0 & 0  \\
\end{tabular}
\end{center}

As a check, it is well-known \cite{barthinvmath} that moduli spaces
of rank 2 bundles on ${\bf P}^2$ of $c_1 = 0, c_2 = n$ have dimension
$4n-3$.  In this case, ${\cal E}$ has $c_1 = 0, c_2 = 3$,
so it has 9 complex deformations -- 
and indeed, $\mbox{dim Ext}^1_{{\bf P}^2}({\cal E},{\cal E}) = 9$.

This result for $\mbox{Ext}^1$ also implies that ${\cal E}$ has
no ``equivariant'' deformations -- that is, deformations that
preserve equivariance.  (In general, $\mbox{dim Ext}^1({\cal E}, {\cal E})_{
\chi = 0}$ is the number of sheaf moduli that preserve equivariance.) 
This also is relatively straightforward to check.

The groups $\mbox{Ext}^i_{{\bf P}^2}({\cal E},{\cal E}\otimes {\cal O}(K))_{\chi}$
can either be worked out directly or calculated via equivariant Serre duality.

Putting those two sets of $\mbox{Ext}$ groups together, we find
that
$\mbox{Ext}^0_{E}({\cal E},{\cal E})_{\chi}$ is given by
\begin{center}
\begin{tabular}{c|cccccc}
 & -2 & -1 & 0 & 1 & 2 & 3 \\ \hline
3 & 0 & 0 & 0 & 0 & 0 & 0 \\
2 & 0 & ${\bf C}$ & 0 & 0 & 0 & 0 \\
1 & 0 & 0 & 0 & 0 & 0 & 0 \\
0 & 0 & 0 & ${\bf C}$ & 0 & 0 & 0 \\
-1 & 0 & ${\bf C}$ & 0 & 0 & ${\bf C}$ & 0 \\
-2 & 0 & 0 & 0 & 0 & 0 & 0 \\
\end{tabular}
\end{center}
and 
\begin{displaymath}
\mbox{Ext}^1_E({\cal E},{\cal E})_{\chi} \: = \:
\mbox{Ext}^0_E({\cal E},{\cal E})_{- \chi}
\end{displaymath}

Our result for $\mbox{Ext}^1_{E}$ reveals some unfortunate
pathologies that must be dealt with.  It is well-known
\cite{atiyah57} that any moduli space of bundles on a smooth
elliptic curve is isomorphic to the elliptic curve itself,
so for smooth $E$, $\mbox{dim Ext}^1_{E}({\cal E},{\cal E}) = 1$.
Yet here, by contrast, $\mbox{dim Ext}^1_{E}({\cal E},{\cal E}) > 1$.
How is this possible?  Strictly speaking $\mbox{Ext}^1({\cal E},{\cal E})$
is the Zariski tangent cone to a moduli space of sheaves at sheaf
${\cal E}$.  If the moduli space is singular at the point represented
by sheaf ${\cal E}$, for example, then $\mbox{Ext}^1({\cal E},{\cal E})$
need not have the same dimension as the moduli space.
In particular, in the present case $E$ is extremely singular,
and so there is no good reason why $\mbox{dim Ext}^1_E({\cal E},{\cal E})
 = 1$.  The ``extra'' generators of $\mbox{Ext}^1$ should be
thought of as closely analogous to ``fake'' marginal operators
in string theory -- that is, operators with the right conformal
dimension to be marginal, but which do not actually represent
flat directions in the theory.

\subsection{Elliptic Curve in ${\bf P}^2 / {\bf Z}_3$ }

\begin{figure}
\centerline{\psfig{file=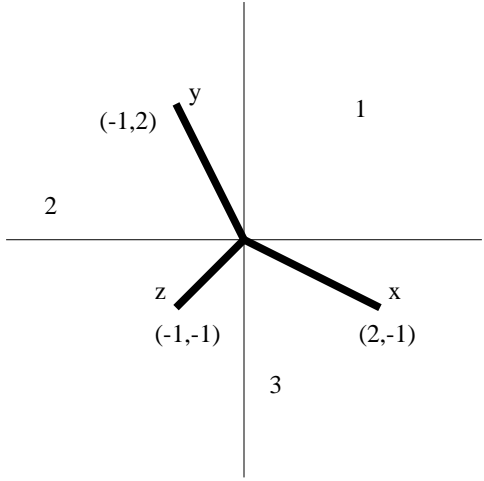,width=2.5in}}
\caption{\label{p2z3} A fan for the toric variety ${\bf P}^2 / {\bf Z}_3$.}
\end{figure}

A fan describing ${\bf P}^2 / {\bf Z}_3$ as a toric variety
is shown in figure~\ref{p2z3}.

Strictly speaking the sheaf cohomology calculations presented
in \cite{meallen} should only be trusted on a smooth toric variety,
however we shall apply them here.

The calculations in this case are extremely similar to those
already presented, so we shall simply summarize the results.

Let ${\cal E}$ be a sheaf on ${\bf P}^2/{\bf Z}_3$ defined by
the same type of filtrations used to define the ${\cal E}$
used in the previous section, then for an elliptic curve $\tilde{E}$
(the mirror to $E$) defined by 
\begin{displaymath}
xyz = 0 
\end{displaymath}
we have
\begin{displaymath}
H^0 ( \tilde{E}, {\cal E})_{\chi} \: = \: H^1 ( \tilde{E}, 
{\cal E} )_{\chi} \: = \:
0 \: \forall \chi
\end{displaymath}
\begin{displaymath}
H^0 ( \tilde{E}, {\cal E}^{\vee})_{\chi} \: = \: H^1 ( \tilde{E}, 
{\cal E}^{\vee} )_{\chi} 
\: = \:
0 \: \forall \chi
\end{displaymath}
and $\mbox{Ext}^1_{\tilde{E}}({\cal E},{\cal E})_{\chi}$
is given by
\begin{center}
\begin{tabular}{c|ccccc}
 & -2 & -1 & 0 & 1 & 2 \\ \hline
2 & 0 & 0 & 0 & 0 & 0 \\
1 & 0 & 0 & 0 & ${\bf C}$ & 0 \\
0 & 0 & ${\bf C}$ & ${\bf C}$ & 0 & 0 \\
-1 & 0 & 0 & ${\bf C}$ & 0 & 0 \\
-2 & 0 & 0 & 0 & 0 & 0 \\
\end{tabular}
\end{center}
and
\begin{displaymath}
\mbox{Ext}^0_{\tilde{E}}({\cal E},{\cal E})_{\chi} \: = \:
\mbox{Ext}^1_{\tilde{E}}({\cal E},{\cal E})_{- \chi}
\end{displaymath}

Here again we have the unfortunate phenomenon that there are
more generators in $\mbox{Ext}^1_{\tilde{E}}$ than there should be 
genuine moduli.

How do these results compare to those for the mirror $E$?
As only one of the groups ($\mbox{Ext}^1({\cal E},{\cal E})$)
is nonzero, and $\mbox{dim Ext}^1_{E}({\cal E},{\cal E}) =
\mbox{dim Ext}^1_{\tilde{E}}({\cal E},{\cal E})$,
there exists an isomorphism of isotypic decompositions,
and so this agrees -- trivially -- with the conjecture.

\section{Higher Dimensional Calabi-Yaus}

In principle one could work through the same calculations
for higher dimensional Calabi-Yaus, though in practice these
computations are more cumbersome.  We shall not do so here,
but instead will mention a few subtleties associated with such
computations.

The tangent bundle of the equivariant elliptic curve was the
restriction of a bundle on ${\bf P}^2$, but in higher dimensions
the tangent sheaf of an equivariant Calabi-Yau will instead be
the restriction of an equivariant torsion-free sheaf on the 
ambient toric variety.  

As a result, instead of calculating sheaf cohomology groups
to count complex and Kahler moduli, one should consider
Ext groups.  More precisely, $H^1(X, T^* X)$ is replaced
with $\mbox{Ext}^1_X(TX, {\cal O})$.  The group $H^1(X, TX)$
is always isomorphic to $\mbox{Ext}^1_X({\cal O}, TX)$.

Global Ext groups of equivariant sheaves have an isotypic decomposition,
just as the sheaf cohomology groups, and are relatively
straightforward to calculate.  As the methods involved
may not be familiar to the reader, we review a simple example
in appendix~\ref{idealap}.

\section{Conclusions}

In this paper we have presented a conjecture regarding how (0,2)
mirror symmetry acts on sheaf deformations.  More precisely,
we have made a conjecture concerning how deformations of the
tangent bundle of mirror symmetric Calabi-Yaus are mirror mapped
into each other, in a limit that all worldsheet instanton
corrections on both sides of the mirror are turned off.

We would like to make a more general conjecture concerning
mirror symmetry.  For Calabi-Yaus realized as hypersurfaces
in toric varieties, an ``inherently toric'' description of
mirror symmetry has been well-developed -- Kahler moduli descending
from the ambient space are exchanged with complex structure deformations
to other hypersurfaces in the same ambient space. 
For Calabi-Yaus realized as special Lagrangian fibrations \cite{syz},
mirror symmetry maps such Calabi-Yaus into other Calabi-Yaus also
realized as special Lagrangian fibrations.  More generally,
it would appear that for Calabi-Yaus with any defining property ``X'',
there exists a description of mirror symmetry that is inherently
``X''-ic.  Perhaps such an observation can be made more precise -- 
perhaps there exists some universal (in the sense of category
theory) definition of mirror symmetry, which yields specific
mirror conjectures for any given family of Calabi-Yaus.

\section{Acknowledgements}

We would like to thank P. Aspinwall, A. Knutson, and E. Witten 
for useful conversations.

\appendix

\section{Complex structure moduli spaces}  \label{compmodspace}

Moduli spaces of complex structures can be constructed
as GIT quotients \cite{git,newstead} (see the appendix of
\cite{meallen} for an introduction to GIT quotients for
physicists).  Typically smooth
hypersurfaces are stable, and singular hypersurfaces are
either semistable or (more commonly) unstable.

In this appendix we shall examine moduli spaces of complex
structures explicitly, and check that the degenerate
hypersurfaces used in the text really are on the moduli space
(i.e., are semistable).  Our presentation will be closely
analogous to those in \cite[section 4.2]{git} and 
\cite[section 4.4]{newstead}.

Let $x_0$, $x_1$, \ldots $x_n$ be homogeneous coordinates
on ${\bf P}^n$.  We shall study moduli spaces of degree $n+1$
hypersurfaces in ${\bf P}^n$.  

To set notation, let 
\begin{displaymath}
\sum \, a_{i_1 i_2 \cdots i_n} x_0^{n+1 - i_1 - i_2 - \cdots - i_n} 
x_1^{i_1} x_2^{i_2}
\cdots x_n^{i_n} \: = \: 0
\end{displaymath}
define a degree $n+1$ hypersurface in ${\bf P}^n$.

Now, not every distinct set of $a_{i_1 i_2 \cdots i_n}$ defines
a distinct hypersurface -- hypersurfaces related by 
an action of $GL(n+1,{\bf C})$ on the homogeneous coordinates
are identical.  As the overall breathing mode of $GL(n+1,{\bf C})$
is irrelevant, it suffices to consider the action of
$SL(n+1,{\bf C})$.

The moduli space of complex structures is simply a GIT quotient
of the space of $a_{i_1 i_2 \cdots i_n}$ (namely,
${\bf C}^k$ for some $k$) by $SL(n+1,{\bf C})$ (with ample
line bundle ${\cal O}$ on ${\bf C}^k$).

How can we determine whether a hypersurface is stable, semistable,
or unstable under this GIT quotient?  We shall use the numerical
criterion for stability (see \cite[section 2.1]{git} or 
\cite[section 4.2]{newstead} or the appendix to \cite{meallen}).
This involves checking stability under all one-parameter subgroups
of $SL(n+1,{\bf C})$. 

Now, every one-parameter subgroup of $SL(n+1,{\bf C})$ is
conjugate to one of the form
\begin{displaymath}
\lambda(t) \: = \: \mbox{diag}\left( t^{r_0}, t^{r_1}, \cdots 
t^{r_n} \right)
\end{displaymath}
with
\begin{displaymath}
\sum r_i \: = \: 0
\end{displaymath}
and
\begin{displaymath}
r_0 \: \geq \: r_1 \: \geq \: \cdots \: \geq \: r_n
\end{displaymath}
with not all $r_i$ zero.
Clearly, such a one-parameter subgroup will act on the
$a_{i_1 i_2 \cdots i_n}$ as
\begin{displaymath}
\lambda(t) \, a_{i_1 i_2 \cdots i_n} \: = \: 
t^{ - r_0 (n+1 - i_1 - i_2 - \cdots - i_n) - r_1 i_1 - \cdots - 
i_n r_n} \, a_{i_1 i_2 \cdots i_n}
\end{displaymath}

Now, define
\begin{eqnarray*}
\mu( \{ a_{i_1 i_2 \cdots i_n} \} , \lambda) & = & 
\mbox{unique integer } \mu \mbox{ such that }
                      \lim_{t \rightarrow 0} t^{\mu} \lambda(t) \, a_{i_1
                      i_2 \cdots i_n} 
                      \mbox{ exists and is nonzero} \\
                & = & \mbox{max } \{ r_0 (n+1 - i_1 - i_2 - \cdots
       - i_n) + r_1 i_1 + \cdots + i_n r_n \, | \, a_{i_1 i_2 \cdots
i_n} \neq 0 \}
\end{eqnarray*}
According to the numerical criterion for stability,
a point will be stable precisely when $\mu > 0$ for all one-parameter
subgroups $\lambda$, semistable when $\mu \geq 0$ for all one-parameter
subgroups $\lambda$, and unstable otherwise. 

It is now easy to check that the equivariant Calabi-Yau defined
by the hypersurface
\begin{displaymath}
x_0 x_1 \cdots x_n \: = \: 0
\end{displaymath}
in ${\bf P}^n$ is always properly semistable.  For this hypersurface,
for any one-parameter subgroup $\lambda(t)$,
\begin{eqnarray*}
\mu & = & r_0 ( 1 ) \: + \: r_1 \: + \: \cdots \: + \: r_n \\
    & = & 0 
\end{eqnarray*}
Thus, equivariant Calabi-Yaus of this form are always semistable,
and more to the point always exist on the moduli space.

Now we shall specialize to the case of elliptic curves in ${\bf P}^2$.
Suppose the elliptic curve has a singular point, and that
(without loss of generality) the point is located at
\begin{displaymath}
(x_0, x_1, x_2) \: = \: (1, 0, 0)
\end{displaymath}
In this case, we have constraints on the \{ $a_{i_1 i_2} \}$:
$a_{i_1 i_2} = 0$ if $i_1 = i_2 = 0$,
or if precisely one of the $i_j = 1$ and the other is zero.
We claim that in this case, there exists a one-parameter
subgroup $\lambda$ such that $\mu(a_{i_1 i_2}, \lambda) \leq 0$.
Let $r_0 = 2$, $r_1 = r_2 = -1$, then
\begin{eqnarray*}
\mu( \{ a_{i_1 i_2} \}, \lambda) & = & 2 (3 - i_1 - i_2) - i_1 - i_2 \\
 & = & 3 ( 2 - i_1 - i_2 ) \\           
 & \leq & 0 \: \mbox{ as } i_1 + i_2 \geq 2 \mbox{ for } a_{i_1 i_2} \neq 0
\end{eqnarray*}
Thus, a singular elliptic curve in ${\bf P}^2$ can never be
properly stable, but is either properly semistable or unstable.

\section{Ideal sheaf on ${\bf P}^2$}    \label{idealap}

In order to gain some basic experience with 
nontrivial global $\mbox{Ext}$ calculations, we shall warm up
by studying an ideal sheaf ${\cal I}$ on ${\bf P}^2$, vanishing
to order 1 at $x = y = 0$.  (For notation, consult figure~\ref{p2}.)
In particular, we shall calculate the global Ext groups
$\mbox{Ext}^0({\cal I},{\cal O})$ and $\mbox{Ext}^1({\cal I},{\cal O})$.

What modules do we associate to cones to describe this ideal
sheaf ${\cal I}$?  Over all cones $\sigma$ except the dimension 2 cone spanned
by $x$, $y$, the module associated to $\mbox{Spec } {\bf C}[
\sigma^{\vee}]$ is precisely ${\bf C}[\sigma^{\vee}]$.
Over cone 1, spanned by $x$, $y$, the module is the ideal
$I = (x,y) \subset {\bf C}[x,y]$.

Now, let us calculate global Ext groups of this ideal sheaf,
following the prescription in \cite{meallen}.
Let $I^{\sigma}$ denote
the module defining ${\cal I}$ on cone $\sigma$,
${\cal O}$ the structure sheaf, and ${\cal O}^{\sigma}$
the module defining ${\cal O}$ on cone $\sigma$
 -- namely, ${\cal O}^{\sigma} = {\bf C}[\sigma^{\vee}]$.
In this notation, the global Ext group $\mbox{Ext}^n({\cal I}, 
{\cal O})_{\chi}$ is the limit of a spectral
sequence with first-level terms
\begin{displaymath}
E_1^{p,q} \: = \: \bigoplus_{codim \: \sigma = p} \, 
\mbox{Ext}^{q}_{ {\bf C}[\sigma^{\vee}]}(I^{\sigma}, 
{\cal O}^{\sigma})_{\chi}
\end{displaymath}
where the Ext groups on the right side are the usual 
Ext groups of modules.  It is straightforward to calculate
these, with the results
\begin{eqnarray*}
\mbox{Ext}^0 ( {\bf C}[\sigma^{\vee}], {\bf C}[\sigma^{\vee}] )
& = & {\bf C}[ \sigma^{\vee}] \\
\mbox{Ext}^n ( {\bf C}[\sigma^{\vee}], {\bf C}[\sigma^{\vee}]  )
& = & 0 \mbox{ for } n > 0 \\
\mbox{Ext}^0 ( I, {\bf C}[x,y] ) & = & {\bf C}[x,y] \\
\mbox{Ext}^1 ( I, {\bf C}[x,y] ) & = & {\bf C}[x,y] / I \\
\mbox{Ext}^n ( I, {\bf C}[x,y] ) & = & 0 \: \mbox{ for } n > 1
\end{eqnarray*}

Now, let us calculate the second-level terms.
Recall
\begin{displaymath}
d_r: E_r^{p,q} \: \rightarrow \: E_r^{p+r, q-r+1}
\end{displaymath}
so in particular
\begin{displaymath}
d_1: E_1^{p,q} \: \rightarrow \: E_1^{p+1, q}
\end{displaymath}
and this differential is precisely the \u{C}ech differential.
Thus, $E_2^{n,0} = H^n ( {\it Hom}({\cal I}, {\cal O} ) )$,
and ${\it Hom}({\cal I}, {\cal O}) = {\cal O}$, so we can
read off
\begin{eqnarray*}
E_2^{0,0} & = & \left\{ \begin{array}{ll}
                {\bf C} & \chi = 0 \\
                0       & \mbox{otherwise}
                \end{array} \right.  \\
E_2^{1,0} & = & 0 \: \forall \chi \\
E_2^{2,0} & = & 0 \: \forall \chi 
\end{eqnarray*}
In principle, identical methods give $E_2^{n,1} = H^n( {\it Ext}^1
( {\cal I}, {\cal O} ) )$.  However, ${\it Ext}^1( {\cal I},
{\cal O}  )$ is not torsion-free, so we should go over
the calculation in somewhat more detail.

Even for an equivariant torsion sheaf, we can still
calculate sheaf cohomology (on the obvious Leray cover)
as \u{C}ech cohomology of the complex
\begin{displaymath}
0 \: \rightarrow \: \bigoplus_{codim \: \sigma = 0} E^{\sigma}(\chi)
\: \rightarrow \: \bigoplus_{codim \: \sigma = 1} E^{\sigma}(\chi)
\: \rightarrow \: E \: \rightarrow \: 0
\end{displaymath}
where $E^{\sigma}$ is the module defining the sheaf
${\it Ext}^1 ( {\cal I}, {\cal O})$ over the open toric neighborhood
associated to cone $\sigma$, namely
\begin{displaymath}
E^{\sigma} \: = \: {\it Ext}^1 ( I^{\sigma}, {\cal O}^{\sigma})
\end{displaymath}
Over the open torus orbit, the equivariant sheaf
${\it Ext}^1( {\cal I}, {\cal O})$ is a trivial rank 0 vector
bundle, so $E = 0$.  Also, from results above,
$E^{\sigma} = 0$ for $\sigma$ a cone of codimension 1.
Thus, the cohomology of this complex is trivial to compute,
and we find
\begin{eqnarray*}
E_2^{0,1} & = & {\bf C}[x,y]/I \: \cong \: \left\{  \begin{array}{ll}
                 {\bf C} & \chi = 0 \\
                 0       & \mbox{otherwise}
                 \end{array} \right.  \\
E_2^{1,1} & = & 0 \: \forall \chi \\
E_2^{2,1} & = & 0 \: \forall \chi
\end{eqnarray*}

Now, it is easy to check that
\begin{eqnarray*}
E_{\infty}^{0,0} & = & E_2^{0,0} \\
E_{\infty}^{1,0} & = & E_2^{1,0} \\
E_{\infty}^{0,1} & = & E_2^{0,1}
\end{eqnarray*}
so as a result we have
\begin{eqnarray*}
\mbox{Ext}^0( {\cal I}, {\cal O} )_{\chi} & = & E_2^{0,0} \: = \:
\left\{ \begin{array}{ll}
{\bf C} & \chi = 0 \\
0       & \mbox{otherwise} 
\end{array} \right. \\
\mbox{Ext}^1( {\cal I}, {\cal O} )_{\chi} & = & E_2^{1,0} \oplus
E_2^{0,1} \: = \:  \left\{ \begin{array}{ll}
{\bf C} & \chi = 0 \\
0       & \mbox{otherwise} 
\end{array} \right.
\end{eqnarray*}
which agrees precisely with the known (in fact, standard) result.

\end{document}